\documentclass[%
 reprint,
nofootinbib,
 amsmath,amssymb,
 aps,
]{revtex4-2}

\usepackage{hyperref}
\usepackage{url}
\usepackage{graphicx}% Include figure files
\usepackage{dcolumn}% Align table columns on decimal point
\usepackage{bm}% bold math
\begin{document}

% \preprint{APS/123-QED} %TODO

\title{Cosmic Piano:\\analysing the sound of the Universe}% Force line breaks with \\

\author{ Guillermo Tejeda Muñoz$^1$, Luis Alberto Perez Moreno$^2$, Simone Ragoni$^3$, Hector David Regules Medel$^1$, Arturo Fernández Téllez$^1$, Yael Antonio Vasquez Beltran$^1$\\}
 % \altaffiliation[Also at ]{Physics Department, XYZ University.}
\affiliation{%
 $^1$Benemerita Universidad Autonoma de Puebla, Mexico\\
 $^2$Princeton University, USA\\
 $^3$Creighton University, USA\\
\\
}%

% \author{Ann Author}
%  \altaffiliation[Also at ]{Physics Department, XYZ University.}%Lines break automatically or can be forced with \\
%  \email{Second.Author@institution.edu}
% \affiliation{%
%  Authors' institution and/or address\\
%  This line break forced with \textbackslash\textbackslash
% }%

% \collaboration{MUSO Collaboration}%\noaffiliation

%  \homepage{http://www.Second.institution.edu/~Charlie.Author}
% \affiliation{
%  Second institution and/or address\\
%  This line break forced% with \\
% }%
% \affiliation{
%  Third institution, the second for Charlie Author
% \affiliation{%
%  Authors' institution and/or address\\
%  This line break forced with \textbackslash\textbackslash
% }%

% \collaboration{CLEO Collaboration}%\noaffiliation

\date{\today}% It is always \today, today,

\begin{abstract}
% We present a new complete Masterclass opportunity for outreach to the general public providing a unique and complete experience, from the design, to the construction of the experiment, to analysing the data. The Cosmic Piano is a set of scintillator-based modules with avalanche photodiodes (APD) readout, designed to detect the muons from the air showers generated by the impact of the Cosmic Rays with the atmosphere. A note and a flash of light are produced whenever the crossing of a muon is signalled in one of the modules of the Cosmic Piano, and the events from each of them are stored together with the timestamp of the event, allowing for further offline analysis for the Masterclass. 

We present the Cosmic Piano, an innovative outreach tool designed to detect muons from air showers produced by cosmic rays colliding with the atmosphere. The Cosmic Piano consists of scintillator-based modules with avalanche photodiodes (APD) readout, which produce a musical note and a flash of light whenever a muon passes through one of them and triggers its electronics. The Cosmic Piano has been showcased at various high-profile events, including the Montreux Jazz Festival and the EuroScience Open Forum, seamlessly fusing scientific phenomena with live music. % During these performances, the random intervals of cosmic rays create a unique canvas for musicians to improvise in real-time, making the universe a part of the musical composition. 
Building on its success, we have developed a new, comprehensive Masterclass session aimed at engaging high school and university students. This session includes analysing both simulated cosmic ray events and real data from the Cosmic Piano. The Masterclass provides participants with a hands-on experience that mirrors the experimental procedures used in major physics experiments, such as those conducted at CERN with the Large Hadron Collider (LHC) collision data. This initiative offers a unique opportunity for students to engage with an active experiment, bridging the gap between outreach and serious scientific research.

\end{abstract}

\maketitle

%\tableofcontents

\section{\label{sec:intro}A Piano for Cosmic Rays}

Cosmic rays are an abundant and natural source of particles, which have been the object of particle and astroparticle physics investigations since their discovery by Victor Hess in 1912 in balloon experiments \cite{Hess:1912srp}. After more than a hundred years from their discovery, Cosmic Rays have been used in multiple contexts, with several different applications. Astroparticle physicists are for example interested in studying the composition of the Cosmic Rays, and thus they have since then built vast arrays to detect air showers, i.e. the interaction of said Cosmic Rays with the atmosphere. Collider Physicists working on collider experiments, e.g. the ALICE and CMS experiments at the CERN LHC, typically use Cosmic Rays to study and perform the alignment of their detectors \cite{ALICE:2010tia}. Nowadays Cosmic Rays are of great interest for their applications to Muon Tomography, and thus their usage to effect the muon analogue to an \textit{x-ray} of e.g. the magmatic chamber of existing volcanoes, vehicles, or even more recently there was a fascinating example of using Muon Tomography to study the inside of the Giza Pyramid, which led to the discovery of a hidden tunnel \cite{morishima2017discovery}. Extensive reviews of the Physics of Cosmic Rays and their applications can be found online (one notable example is \cite{gaisser1990cosmic}) and are not the aim of the current contribution.

Cosmic Rays have also been used to show the general public the daily life of particle and astroparticle physicists, thus complementing already existing particle physics outreach programs, within the context of the International Masterclasses organised by the IPPOG (International Particle Physics Outreach Group) Collaboration \cite{ippog-imc, cecire2018ippog}. One such an example is the Masterclass organised by the Pierre Auger Collaboration \cite{pierreauger:masterclasses}. 

The contribution here presented aims to show a new complete Masterclass program, which brings the audience through the complete journey of the particle physicist, from designing, to building a particle detector, to analysing the data collected with said experiment. This was usually the domain of just the laboratory classes for nuclear and particle physicists, where the students learn to take their first steps in their journeys as experimental physicists. The only slightly comparable experience is the construction of cloud chambers, which can be done for example at the new CERN Science Gateway \cite{cern:sciencegateway}, but is otherwise challenging to implement outside of specially constructed areas, due to the complexity of the detector itself (stable operations of a cloud chambers are not possible without dedicated machineries).

To overcome these shortcomings of real life particle physics experiments for outreach purposes, we have in fact built a Cosmic Piano\footnote{WIPO Patent No. WO2015184791A1. Available: \url{https://patentscope.wipo.int/}},%\footnotetext[2]{Here's the second.}
a scintillator-based detector with avalanche photodiode readout, which plays a musical note when a cosmic muon passes through the acceptance of the detector. The Cosmic Piano has been shown to have powerful outreach capabilities, as already displayed in several occasions such as very recently at the International Conference in High Energy Physics (ICHEP) 2024. Owing to its simplicity, we have decided to increase the coverage of the Cosmic Piano, by also implementing a Masterclass, which targets high school and university students by making use of an intuitive programming approach with Jupyter Notebooks written in Python \cite{python3}.
 
\section{\label{sec:detector}How to build a Cosmic Piano}

\textbf{Detector components} \\

\begin{figure}[h]
\includegraphics[width=0.95\columnwidth]{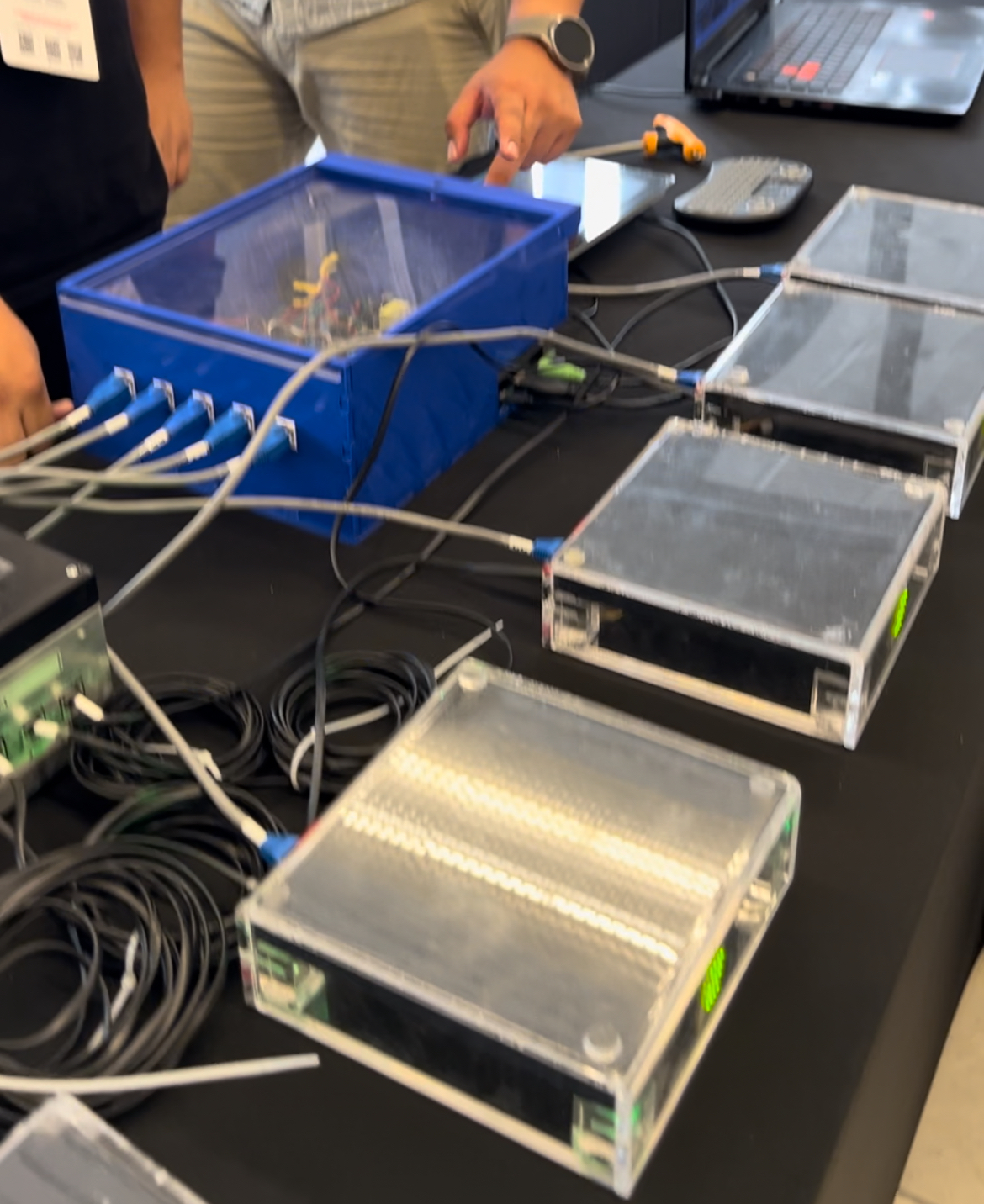}% Here is how to import EPS art
\caption{\label{fig:cosmicpiano} The Cosmic Piano being shown at ICHEP 2024. The typical setup of the Cosmic Piano is with five independent modules, which are readout by shared electronics. An audio system is linked to the Raspberry Pi in the final stage, and a musical note is played by it whenever the passage of a muon is detected.}
\end{figure}

The Cosmic Piano is an ensemble of a certain number of independently operating modules, typically five of them, as shown in Fig.~\ref{fig:cosmicpiano}.

%Each of the modules of the Cosmic Piano is essentially made up of single slab of scintillator, of sizes 14x14x1 cm$^3$, which is readout by avalanche photodiodes (APDs), and amplified. 

Each module consists of a slab of scintillating plastic measuring 14x14x1 cm$^3$. This material has the unique property of emitting light (photons) when a particle passes through it. Photons generated by the scintillating plastic are collected by a phase shift fibre embedded within the plastic. The collected photons travel through the fibre to the ends where they are detected by a pair of avalanche photodiodes (APDs) \cite{hamamatsu:apd}. APDs utilize the well-known photoelectric effect, when photons impact the electrons in the metal of the APD, they cause the electrons to move, thereby generating an electrical signal. The electrical signals generated by the APDs in each module are transmitted to the main electronic system, which is responsible for conditioning and processing these signals. The electronic system is composed of a custom-made electronic board, a Field Programmable Gate Array (FPGA), a Raspberry Pi, and a touchscreen. Each module has Light Emitting Diodes (LEDs) in its corners that light up every time they detect a particle, providing a visual indication of particle detection. \\

\begin{figure}[h]
\includegraphics[width=0.60\columnwidth]{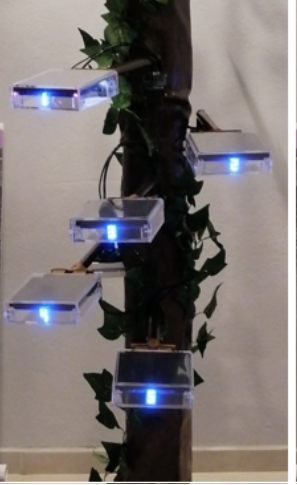}
\caption{\label{fig:cosmictree} Cosmic Tree arrangement. The \textit{Cosmic Tree} is the mechanical structure utilized to mount the Cosmic Piano modules. In this arrangement, each branch of the Cosmic Tree corresponds to a different module of the Cosmic Piano. This arrangement allows flexible configuration and precise detection of cosmic events.}
\end{figure}

An additional component of the Cosmic Piano is a mechanical structure designed to mount and align the modules efficiently, ensuring that they are arranged vertically. This facilitates both single-mode and coincidence-mode operations. This configuration, known as the \textit{Cosmic Tree}, is depicted in Fig.~\ref{fig:cosmictree}.

\textbf{Electronics system} \\

\begin{figure}[h]
\includegraphics[width=0.95\columnwidth]{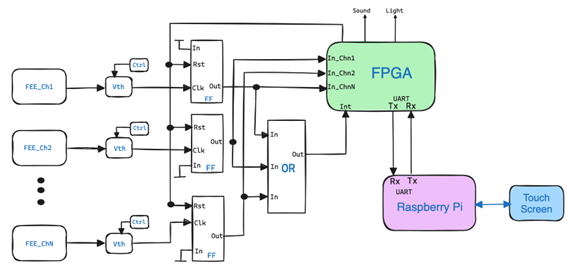}
\caption{\label{fig:electronicsdiagram} Diagram of the electronics. The diagram illustrates the key components of the electronic system, including the discrimination stage, trigger generation logic stage, FPGA, Raspberry Pi, and touchscreen interface.}
\end{figure}

The electronics system, as illustrated in Fig.~\ref{fig:electronicsdiagram}, is responsible for receiving and processing the signals detected by the Cosmic Piano modules. 

The process begins with the discrimination stage, where the signals are amplified, filtered and compared against a predefined threshold to ensure that only significant events are processed. After this initial filtering, the signals move to the trigger generation logic stage.

Next, the signals are processed by a Field Programmable Gate Array (FPGA). The FPGA plays a crucial role in configuring the system’s operational parameters and managing the flow of data based on the programmed logic.

Finally, the processed data are sent to a Raspberry Pi 4, which handles communication with a touchscreen interface and a pair of speakers, connected to its ports. The audio and visualization system is obtained through the Raspberry Pi 4. Additionally, it allows users to set and adjust configurations for the Cosmic Piano. The signals are correlated with a timestamp from the Raspberry Pi, with a precision of 1 $\mu$s, and stored together with the identifier of the module that triggered the event for the data analysis part of the Masterclass.

%The signals from each module are then sent to an FPGA, for further analysis and processing. Here is the signal is evaluated and sent to a Raspberry Pi. The signals are correlated with a timestamp from the Raspberry Pi, with a precision of 1 $\mu$s, and stored together with the identifier of the module that triggered the event for the data analysis part of the Masterclass.

\textbf{FPGA logic and user interface} \\

The Cosmic Piano operates by linking each detection module to a specific musical note. When a particle passes through a module, it triggers the associated note, which is played through a pair of speakers connected to the Raspberry Pi 4. Simultaneously, a flash of light is activated by LEDs installed inside each module. This visual indicator provides immediate feedback whenever a particle is detected.

Additionally, LEDs mounted on the branches of the Cosmic Tree produce flashes according to the configurations set by the user. The illumination of the branches of the Cosmic Tree visually indicates the status and activity of the modules, thereby enhancing the system's interactive and informative capabilities.

To manage and customize the system settings, a small keyboard is located at the bottom of the monitor. This keyboard allows the user to input commands and adjust configurations easily.

When the Cosmic Tree is powered on, a user interface window will appear on the monitor within seconds. This interface displays icons representing the particle counts for each module, providing a real-time view of particle detection. The window also includes configuration options at each end, allowing the operator to enable or disable specific modules as needed. This flexibility ensures that the system can be tailored to various experimental requirements and operational modes. 

Fig.~\ref{fig:userinterface} illustrates the control interface for the Cosmic Piano system. Below is a description of the various screen fields within this interface:

\begin{figure}[h]
\includegraphics[width=0.95\columnwidth]{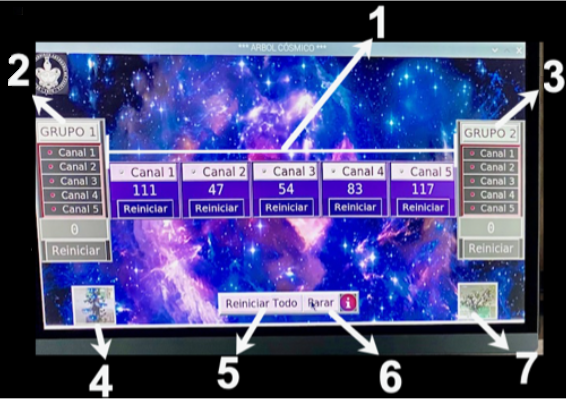}% Here is how to import EPS art
\caption{\label{fig:userinterface} Cosmic Piano User Interface. The user interface provides a comprehensive and interactive platform for real-time control and monitoring. It features various screen fields that allow users to program and adjust settings dynamically.}
\end{figure}

\begin{enumerate}
    \item Module Count fields: Each of them displays the particle count for the corresponding detection module, with Channel 1 representing Module 1, Channel 2 representing Module 2, and so on up to Module 5. At the top of these fields, there is a checkbox that allows users to enable or disable detection for each module with a single click. When a module is disabled, the colour of its field changes to a gray tone, and the audio associated with it is muted, although the particle count continues. At the bottom of each field, there is a reset button that sets the particle count for that specific module back to zero.
    \item Group 1 matching: It allows the user to enable only the selected modules to perform coincidence detection between them. The user can select two or more modules to be included in the matching process. 
    \item Group 2 matching: Same as Group 1, but it allows for a second group to be selected.
    % \item Selective Module Matching field: It allows the user to enable only the selected modules to perform coincidence detection between them. The user can select two or more modules to be included in the matching process. (Group 1)
    % \item Selective Module Matching field: It allows the user to enable only the selected modules to perform coincidence detection between them. The user can select two or more modules to be included in the matching process. (Group 2)
    \item Module 5 Matching field: It enables the user to configure coincidence detection such that any of the modules from 1 to 4 must also show a match with Module 5 for the output to be considered valid. This ensures that the particle must pass through Module 5 in addition to the selected modules.
    \item Global Reset Icon: This icon resets the particle count to zero across all modules simultaneously.
    \item Standby Mode Icon: It puts the entire Cosmic Piano system into standby mode, temporarily pausing all detection and processing activities.
    \item Cosmic Tree Lighting field: It allows the user to control which of the branches of the Cosmic Tree are allowed to produce flashes. 
    % \item Cosmic Tree Lighting field: It allows the user to control the lighting of the Cosmic Tree branches. The branches can be illuminated based on all the modules or selectively based on user preferences. Clicking this window opens a dropdown menu where the user can choose which modules to activate for lighting the branches.
\end{enumerate}

This interface provides comprehensive control over the operation and monitoring of the Cosmic Piano system, allowing for flexible configuration and real-time adjustments. \\

\textbf{Running modes} \\

The Single Mode is the default mode for the Cosmic Piano. In this configuration, all detection modules are activated individually, meaning each module operates independently of the others. Each module is capable of detecting particles that pass through it and, when a particle is detected, the module emits a distinct tone. This mode is useful for general particle detection, where each module functions on its own without needing to coordinate with other modules. This allows for straightforward monitoring of particle activity across all modules simultaneously.

Whereas Coincidence Mode is used to identify particles that pass through two or more modules within a very brief time interval, typically measured in nanoseconds (ns). In this mode, the system is configured to detect coincidences — situations where a particle traverses multiple modules nearly simultaneously. This is important for distinguishing between events caused by a single particle versus multiple particles detected at different times. To increase the probability of detecting such coincidences, the modules can be positioned close to each other, either directly stacked one on top of another or aligned in close proximity. This setup helps to ensure that a particle passing through one module is likely to also pass through the adjacent modules within the defined time window. Such precise placement improves the accuracy of detecting the particle's trajectory and provides more detailed information about its path through the system.

\section{\label{sec:masterclass}The need for a Masterclass}
Masterclasses are an especially accessible way of bringing particle physics and the role of a particle physicist to the public. A prime example of such activities is the International Masterclasses programme run by the IPPOG Collaboration \cite{ippog-imc}. Every year, approximately 13,000 high school and university students in more than 60 different countries participate in this programme, discovering the beauty of particle physics through direct engagement with real data and cutting-edge research techniques.

The International Masterclasses offer a variety of activities that allow students to explore different aspects of particle physics. These include analysing data from major experiments like ATLAS, CMS, and ALICE at CERN, and the MINERvA neutrino experiment at Fermilab. Additionally, there is a specific Masterclass organised by the Pierre Auger Collaboration \cite{pierreauger:masterclasses}, focusing on the study of cosmic rays and the analysis of data from the Pierre Auger Observatory, which is the world’s largest cosmic ray detector.

The Cosmic Piano Masterclass builds on this tradition by offering an immersive experience that goes beyond traditional outreach. It complements the Pierre Auger Masterclass by providing an even more basic and immediate exploration of cosmic rays. By focusing on the direct detection of cosmic ray particles and the simple verification of their existence, the Cosmic Piano introduces participants to the fundamental concept of invisible particles that constitute matter. This hands-on approach makes complex concepts more tangible, allowing participants to grasp the existence of cosmic particles in a straightforward and engaging manner.

The Masterclass begins with a simple explanation of the main physical processes behind cosmic rays, typically lasting one hour, followed by the assembly of the detector modules. This hands-on assembly demystifies the components and operation of a particle detector. Given the precision required for handling scintillators and APDs, these components are initially presented in isolated, protected states to ensure that the assembly process is smooth and informative. Participants can either assemble the Cosmic Piano themselves or observe the process via a remote broadcast. Once assembled, the Cosmic Piano becomes an interactive tool where the detection of cosmic muons is signalled by the emission of notes and LED flashes, providing an immediate and tangible connection to the invisible world of particle physics.

The Masterclass also introduces participants to data analysis, a cornerstone of modern scientific research. Through a user-friendly Jupyter Notebook interface in Python, participants are guided through the analysis of data collected by the Cosmic Piano. This analysis involves examining the triggers for each event, identifying noisy channels, and understanding the importance of data quality control. Additionally, the Masterclass incorporates a simplified version of the Geisser model for cosmic rays \cite{Guan:2015vja}, enabling participants to calculate the expected muon flux at sea level and compare it with their measured data.

The final outcome of the Cosmic Piano Masterclass is a complete, albeit simplified, measurement of the muon flux using all available detector channels. This exercise not only reinforces the participants' understanding of cosmic ray physics but also provides a rare opportunity to experience the entire workflow of an experimental physicist — from detector construction and data collection to analysis and interpretation. Such a comprehensive approach makes the Cosmic Piano Masterclass an invaluable tool in both education and public outreach, helping to inspire the next generation of scientists.

\section{\label{sec:analysis}Data analysis with the Cosmic Piano}
The data analysis part of the Masterclass is available here \cite{cosmic-code}, with the packages loaded as a requirement to the environment created using Binder 2.0 \cite{jupyter2018binder}. It is then opened as a browser tab, with no prior software installation needed. The total length of the exercise with explanations included is about two hours, to allow the students to implement their own exercises. Several types of data are possible: a pseudorandom generator for the modules with configurable rates is given, which can be run independently from the main Jupyter Notebook, along with real Cosmic Piano data. These data are collected in advance since Cosmic Rays require long data taking conditions to have enough statistics to measure the muon flux with good precision.

\begin{figure}[b]
\includegraphics[width=0.95\columnwidth]{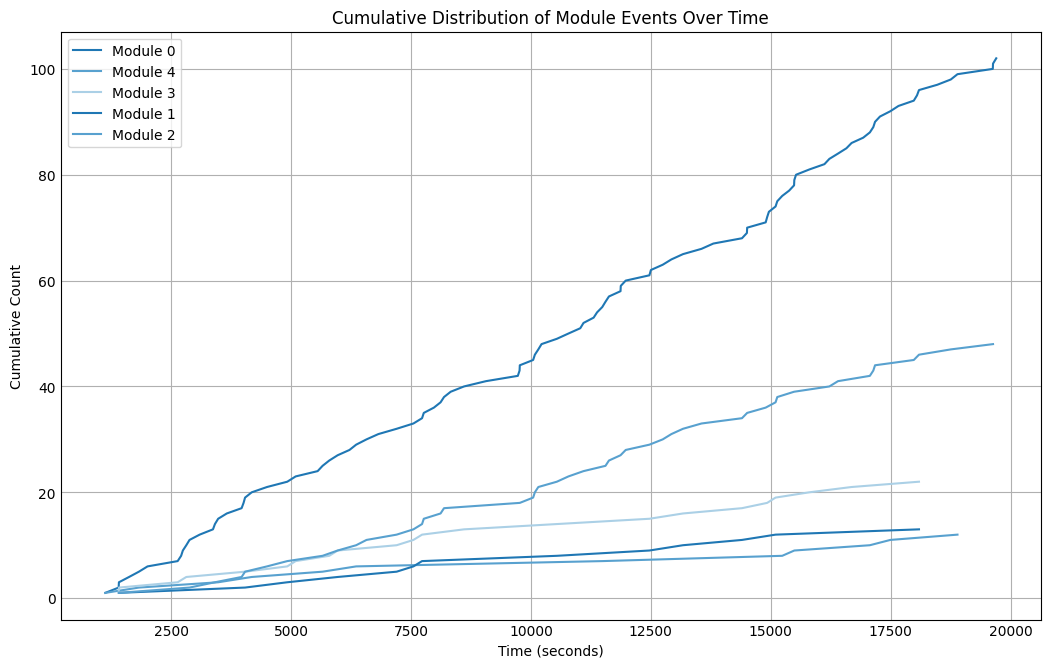}% Here is how to import EPS art
\caption{\label{fig:rates} The rates from the five modules of the Cosmic Piano are shown as a function of time. Since the data taking conditions are similar, i.e. same detectors, close to each other so as not to have different flux conditions, the rates for each of them are expected to be comparable. In case this is not true, it means one or more modules started being noisy, sensitive to background noise.}
\end{figure}

The data analysis starts with quality control of previously collected or randomly generated data. An example is shown in Fig.~\ref{fig:rates}.%
The students learn then about the peculiarities of real life experimental physics, and the joys of running conditions, background noises, and what to expect from healthy data taking conditions.

\begin{figure}[b]
\includegraphics[width=0.95\columnwidth]{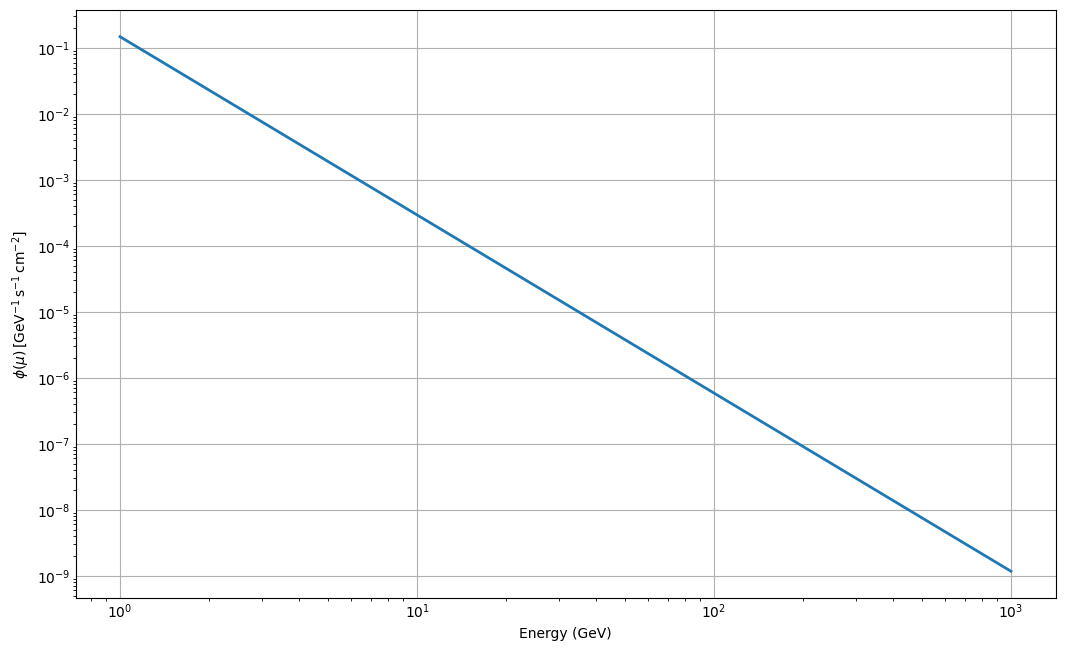}% Here is how to import EPS art
\caption{\label{fig:geissner} A simplified version of the Geisser model for the muon flux. The students will have the chance to think also as a theoretical cosmic ray physicist, by tuning the parameters of the model, to fit their purposes, and observe how the flux varies by changing the inputs to the model.}
\end{figure}

As soon as the quality control phase is terminated, students engage with modelling of the muon flux. We have implemented a simplified version of the Geisser model, which is in line with up-to-date models for muon energies above a few GeV. This is shown in Fig.~\ref{fig:geissner}.%

Finally, the students will be able to measure the muon flux using the Cosmic Piano data, and compare it with the simplified Geisser Model. As the current version of the Cosmic Piano does not allow for the measurement of the energy of the muons, only the integrated flux can be measured. 

Concerning the data taking using the Cosmic Piano, two are the modes which are implemented in the firmware regulating the behaviour of the modules: independent and coincidence mode. In independent mode, the modules operate as independent detectors, each registering the hit of the muons. However, we have decided to also include a coincidence mode, where the detectors are positioned in stacks. In this particular data taking condition, if the modules register a hit within the same time window of 1 $\mu$s, then a coincidence is registered. The reason behind this mode is to illustrate the public the typical behaviour of a muon from the Cosmic Rays, travelling in a straight line, and thus illustrate the nature of a particle, which would be otherwise invisible in our everyday experience.

While this is implemented in firmware, we have also finally implemented coincidence mode also in the offline analysis. In this case, the audience can verify directly using the timestamps of each event, the rate of coincidences. The technique that the audience will have to implement is that of bundling timestamps from the data together, if the timestamps are within 1 $\mu$s distance apart, and counting them. This is particularly instructive, since they can immediately verify their rates with those observed with the real data taking conditions. Furthermore, it is also possible to take the data with different overlaps of the detector modules, and thus also observe the decrease in the rate of the coincidences with the decrease in the overlap of the acceptances of each module.

This coincidence mode brings about one data taking exercise, the \textit{Hodoscope} for vertical muons. To deepen the hands-on experience, students will then conduct this  specific hands-on practice, together with another exercise, which is referred to as the \textit{Extensive Air Shower}. First, they will collect five minutes of cosmic data from two detectors in coincidence mode, starting with a horizontal separation of 1 cm. After each five-minute measurement, the separation will be increased by 4 cm, continuing this process until a separation of 30 cm is reached. The same experiment will then be repeated with the detectors separated vertically. These exercises will then also enable students to understand the results effectively in terms of Extensive Air Shower cosmic events with the horizontal separation, while the Hodoscope exercise further solidifies the notion of the passing of the particles.

\section{\label{sec
}Conclusions}

The Cosmic Piano represents a significant contribution to the realm of scientific outreach and education. This device is not only secure, cost-effective, and easy to operate, but it also requires minimal maintenance, making it an ideal tool for continuous educational use, especially if compared to cloud chambers, which are the only other comparable device. Its modular design allows for flexibility in deployment, whether in a classroom, public science event, or remote learning environment.

A standout feature of the Cosmic Piano is the accompanying Masterclass, which provides a comprehensive, hands-on learning experience that mirrors the procedures used in professional physics experiments, such as those conducted at the experiments operating at the CERN LHC. Participants in the Masterclass are introduced to the full spectrum of particle physics research, from the initial design and construction of the detector to the complex analysis of real and simulated cosmic ray data. The data analysis component of the Masterclass not only teaches important analytical skills but also provides insights into the challenges and rewards of conducting real-world physics experiments. The Cosmic Piano fosters a deeper understanding of particle physics among its participants owing to its simple data analysis interface available through a browser, without prior installation.

Overall, the Cosmic Piano Masterclass offers a unique and powerful educational experience, bridging the gap between scientific outreach and serious research, making cutting-edge physics accessible to a broader audience and inspiring the next generation of scientists.

% \begin{video}
% \href{http://prst-per.aps.org/multimedia/PRSTPER/v4/i1/e010101/e010101_vid1a.mpg}{\includegraphics{vid_1a}}%
%  \quad
% \href{http://prst-per.aps.org/multimedia/PRSTPER/v4/i1/e010101/e010101_vid1b.mpg}{\includegraphics{vid_1b}}
%  \setfloatlink{http://link.aps.org/multimedia/PRSTPER/v4/i1/e010101}%
%  \caption{\label{vid:PRSTPER.4.010101}%
%   Students explain their initial idea about Newton's third law to a teaching assistant. 
%   Clip (a): same force.
%   Clip (b): move backwards.
%  }%
% \end{video}

\begin{acknowledgments}
This work was funded by Consejo Nacional de Humanidades, Ciencias y Tecnologías (CONAHCYT-México) under the project CF-2019/2042, graduated fellowship Grants No. 1071959 and No. 953372, and by the Department of Energy (DOE) of the United States of America (USA) through the grant DE-FG02-96ER40991.
\end{acknowledgments}

\bibliography{apssamp}% Produces the bibliography via BibTeX.

\end{document}